# The XGIS Imaging System on board the THESEUS mission

J.L. Gasent-Blesa*[a], V. Reglero [a], P. Connell [a], B. Pinazo-Herrero [a],
J. Navarro-González [a], P. Rodríguez-Martínez [a], A.J. Castro-Tirado[b,f], M. D. Caballero-García[b],
L. Amati[c], C. Labanti[c], S. Mereghetti[d], F. Frontera[c], R. Campana[c], M. Orlandini[c], J. Stephen[c],
L. Terenzi[c], F. Evangelisti[e], S. Squerzanti[e], M. Melchiorri[e], F. Fuschino[c], A. de Rosa[c], G. Morgante[c].

[a]Image Processing Laboratory, University of Valencia, c/ Catedrático José Beltrán, 2, E46980, Paterna (Valencia), Spain; [b]Instituto de Astrofísica de Andalucía (IAA-CSIC), Glorieta de la Astronomía sn, E18008, Granada, Spain; [c]INAF-OAS Bologna, via P. Gobetti 101, I40129 Bologna, Italy; [d]INAF-IASF Milano, via A. Corti 12, I20133 Milano, Italy; [e]Universitá di Ferrara and INFN (Sezione di Ferrara), via Saragat 1, I44122, Ferrara, Italy; [f]Dpto. de Ingeniería de Sistemas y Automática, Universidad de Málaga, Avda. Cervantes, 2, E29071 Málaga, Spain

## ABSTRACT

Within the scientific goals of the THESEUS ESA/M5 candidate mission, a critical item is a fast (within a few s) and accurate (<15 arcmin) Gamma-Ray Burst and high-energy transient location from a few keV up to hard X-ray energy band. For that purpose, the signal multiplexing based on coded masks is the selected option to achieve this goal. This contribution is implemented by the XGIS Imaging System, based on that technique.

The XGIS Imaging System has the heritage of previous payload developments: LEGRI/*Minisat-01*, *INTEGRAL*, UFFO/*Lomonosov* and ASIM/*ISS*. In particular the XGIS Imaging System is an upgrade of the ASIM system in operation since 2018 on the *International Space Station*. The scientific goal is similar: to detect a gamma-ray transient. But while ASIM focuses on Terrestrial Gamma-ray Flashes, THESEUS aims for the GRBs.

For each of the two XGIS Cameras, the coded mask is located at 630 mm from the detector layer. The coding pattern is implemented in a Tungsten plate (1 mm thickness) providing a good multiplexing capability up to 150 keV. In that way both XGIS detector layers (based on Si and CsI detectors) have imaging capabilities at the medium – hard X-ray domain. This is an improvement achieved during the current THESEUS Phase-A. The mask is mounted on top of a collimator that provides the mechanical assembly support, as well as good cosmic X-ray background shielding.

The XGIS Imaging System preliminary structural and thermal design, and the corresponding analyses, are included in this contribution, as it is a preliminary performance evaluation.

**Keywords:** coded mask, X-ray astronomy, gamma-ray bursts, X-ray transients, imaging techniques, signal multiplexing

## 1. INTRODUCTION

THESEUS (*Transient High Energy Sky and Early Universe Surveyor*) is a European mission candidate developed by a large international collaboration in response to the fifth M-class mission call within the ESA Cosmic Vision program, with a target launch date by 2032 [1, 2].

THESEUS is focused on Gamma-ray Bursts (GRB) and other high-energy transient phenomena over the entirety of cosmic history. Its primary scientific goals address the Early Universe ESA Cosmic Vision themes "How did the Universe originate and what is made of?" and also impact on "The gravitational wave Universe" and "The hot and energetic Universe" fields. More in detail, the main scientific goals of the mission include:

– To explore the Early Universe (cosmic dawn and reionization era) by unveiling a complete census of the Gamma-Ray Burst (GRB) population in the first billion years

*jose.l.gasent@uv.es; phone +34 96 354 3262; uv.es

- To perform an unprecedented deep monitoring of the X-ray transient Universe with extension up to gamma-rays, thus providing a substantial contribution to multi-messenger and time-domain astrophysics, as well as unique and great synergy with the very large observing facilities of the future in both the e.m. (e.g., LSST, ELT, TMT, SKA, CTA, Athena) and multi-messenger domain (advanced second and third Gravitational Waves and neutrino) detectors

These scientific goals are achieved via a unique payload providing a multiwavelength space observatory by means of an unprecedented combination of three main instruments developed by international consortium (mainly Italy, UK, France, Germany, Switzerland, Spain, Belgium, Czech Republic, and Poland) with a wide payload development heritage:

- X-Gamma-rays Imaging Spectrometer (XGIS), which provides a wide and deep sky monitoring in a broad energy band (2 keV-10 MeV) [3]
- Soft X-ray Imager (SXI) that uses lobster-eye telescopes units, allocating capabilities in the soft X-ray band (0.3-5 keV) with a unique combination of high angular resolution [4]
- Infrared Telescope (IRT), a 0.7m near-IR telescope for fast response, with both imaging and spectroscopy capabilities, providing on board near-IR capabilities for immediate transient identification and redshift determination [5]

The mission profile includes a spacecraft (flying in near-equatorial Low-Earth Orbit) with fast (<10°/s) slewing capabilities, a Trigger Broadcasting Unit (TBU, an on-board VHF transmitter) and a THESEUS Burst Alert Ground Segment (TBAGS), i.e. a set of ground VHF antennae located all around the equator.

Figure 1 shows the THESEUS spacecraft concept.

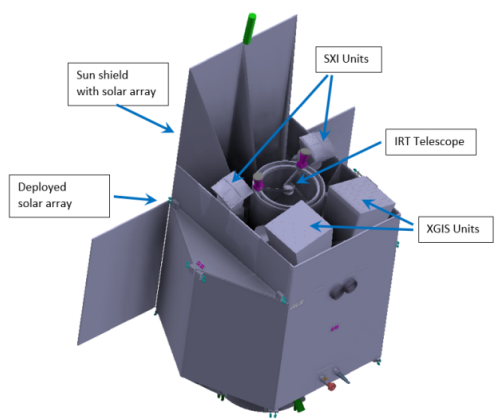

Figure 1 THESEUS spacecraft concept showing the location of the mission payload elements of XGIS, SXI and IRT.

## 2. XGIS INSTRUMENT DESCRIPTION

### 2.1 XGIS instrument overview

Within the THESEUS scientific goals, a fast (within a few s) and accurate (within 15 arcmin) source location in the 2 keV – 150 keV energy band by means of detecting GRBs and X/gamma-ray transients is a critical item [2]. The technique selected is imaging signal multiplexing using a coded mask. The XGIS Imaging System described in this contribution has the heritage of previous payload developments. The XGIS Mask located at 630 mm from the detector layer provides a good multiplexing capability up to 150 keV. Therefore, both XGIS detector layers have imaging capabilities at the X and gamma ray radiation [3, 6].

The X and Gamma-ray Imaging Spectrometer comprises two units (X-gamma ray telescopes) that operate in the range of 2 keV to 10 MeV. The two units are pointed to +/-20° offset directions in such a way that their field of view (FOV) partially overlap.

Each XGIS Unit has imaging capabilities in the low energy band thanks to the combination of a coded mask superimposed to a detector plane. A Passive Shielding placed on the mechanical structure between the Mask and the detector plane determines the FOV for the XGIS imaging up to about 150 keV.

The main elements of the XGIS instrument are: Detector Unit (x2, also referred to as *Cameras*), Data Handling Unit (DHU), XGIS power Supply Units (XSU) and Harness. Each Detector Unit includes a Detection Plane, Electronics (Power Supply and Front-Back electronics), a Coded Mask Assembly and a Collimator Assembly. The XGIS Imaging System includes the Coded Mask Assembly and the Collimator Assembly.

The XGIS Detector Unit (*Camera*) is shown in the following figure (Figure 2).

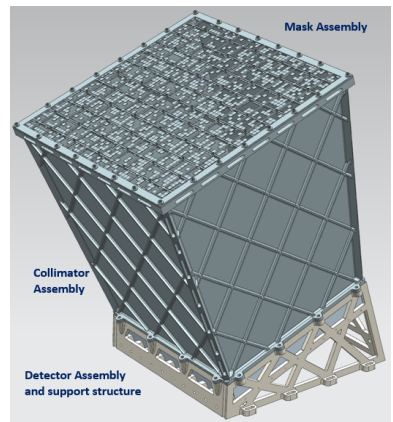

Figure 2 XGIS Detector Unit design concept: Detector Plane plus Electronics, Coded Mask Assembly and Collimator Assembly.

The detection plane of each unit is made of 100 detector modules each one 45x45x63 mm size detecting X-gamma ray in the range of 2 keV – 10 MeV. The position sensitive detector plane is made of 80x80 pixel operating in the 2 keV – 10 MeV range [3]. This large energy range is achieved using low-noise silicon devices, the Silicon Drift Detectors (SDD), both for direct low energy X-radiation, up to 25 – 30 keV, and for the readout of the scintillation light produced by gamma-radiation reaching scintillators crystals beneath the SDD and optically connected to them; a discrimination technique allows to distinguish between X and γ signals.

The double nature of the detector plane leads to a different pixel size for X and γ imaging, the latter extended just up to 150 keV. In the detector design for X-radiation 4800 pixels are 3.5x5 mm in size and 1600 are 3.5x3.5 mm while the 6400 pixels for detection of gamma ray (E > 30 keV) are 4.5x4.5 mm in size with a pitch of 5 mm. The detector plane is organised in 10x10 modules mechanically arranged side by side so than a passive space 5 mm wide is interleaved between one Module and the adjacent ones, in this way there are 9 dead rows and 9 dead columns with width equal to that of one pitch.

## 3. XGIS IMAGING SYSTEM STRUCTURAL REQUIREMENTS

The XGIS Imaging System structural requirements considered in this work come from the project requirements as well as from the heritage of previous missions. These requirements are detailed hereafter:

1) The mass requirement is less than 30 kg, including contingency
2) The coded mask area required is 561x561 mm

3) The envelope requirement is 600x600 mm on top of the Imaging System, and a maximum height of 630 mm (excluding the screw heads)

4) The FOV requirement (for each Detector Unit) is:
   - 77x77 deg in the 2-150 keV energy band
   - 2π sr above 150 keV

5) Stress and Stiffness requirements have been taken from the heritage of previous missions (*INTEGRAL* [7], ASIM/*ISS* [8], *XIPE* [9]), since they have not been defined in the THESEUS project for the period of this work preparation yet. Hence, based on previous developments, the XGIS Imaging System has been conceived to have its fundamental eigenfrequency above 60 Hz, and a quasi-static load for all possible combinations with a maximum acceleration of 24g in each direction. Mission stress and stiffness requirements will be considered in further steps of the project.

## 4. MECHANICAL AND THERMAL DESIGN

### 4.1 XGIS Imaging System envelope and mass

The preliminary design of the XGIS Imaging System described in this work complies with the requirement of the maximum envelope as well as the mass budget (below 30 kg including contingency) detailed in previous section. Figure 3 depicts the geometry of the XGIS Imaging System:

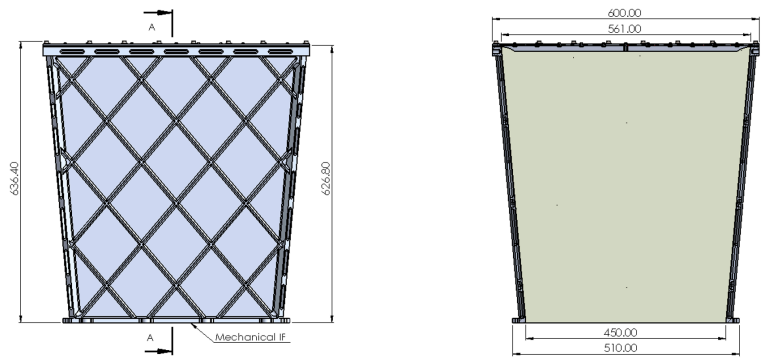

Figure 3 XGIS Imaging System geometrical envelope defined by the Coded Mask Assembly and the Collimator Assembly.

### 4.2 Coded Mask Assembly mechanical design

The XGIS Imaging System has the heritage of previous payload developments: LEGRI/*Minisat-01* [10], *INTEGRAL* [7], UFFO/*Lomonosov* [11] and ASIM/*ISS* [8]. Design concept of the Modular X- and Gamma-Ray Sensor (MXGS) on-board ASIM [12], which is in operation since 2018 at *ISS*, is the most similar one to XGIS among those missions' payloads. The MXGS scientific goal is similar to XGIS one: to detect a gamma-ray transient; Terrestrial Gamma-ray Flashes in ASIM [8, 13], while in THESEUS are the GRBs [1, 2].

Each XGIS Mask Code spatially modulates with transparent and non-transparent pixel elements the incoming X and gamma-ray radiation. The corresponding detection plane detects this Mask-modulated signal.

A random Mask pattern is considered in this preliminary design. The Mask pattern shadows on the detector plane for a given X-gamma ray source located within the XGIS FOV. The image reconstruction is based on a correlation procedure between the detected image and a decoding array from the Mask pattern.

The Coded Mask Assembly envelope is 600x600 mm and has a pattern allowing self-support to guarantee the maximum transparency of the open elements. The Coded Mask Assembly includes the following parts:

- Mask Code
- Mask Support Structure (upper grid, lower grid, and frame)

Figure 4 displays the XGIS Coded Mask Assembly parts:

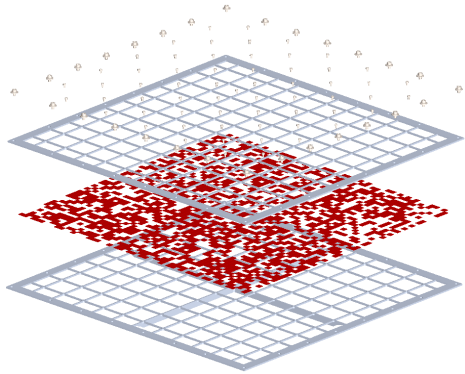

Figure 4 XGIS Coded Mask Assembly exploded view depicting the Mask Code (red colour) and the Mask Support Structure.

The Mask Code of each XGIS unit is made of Tungsten (W) of 1.0 mm thick for the non-transparent pixels; the Mask Code is placed 630 mm above the detector plane (3.2 mm below the Collimator Assembly mechanical interface). The coded area of the Mask is 561x561mm with a pixel size of 10x10 mm square. A random pattern has been considered for this preliminary design.

The Mask Support Structure provides mechanical support to the Code as well as mechanical interface with the Collimator Assembly. This Support Structure have upper grid and a lower grid made of Aluminium alloy that encapsulate the pixels of the Mask Code.

The XGIS imager locates sources via a 50% open, random pattern, Mask Code placed as mentioned at a height 630 mm above the surface of its detector plane. An X-gamma ray source in the FOV of XGIS casts an illumination pattern of the Mask hole pattern on the detector plane, whose (x,y) Centre Point varies with the FOV angular location of the source.

If a source is observed at some FOV zenith/azimuth angle (z,a) relative to the detector normal and (x,y) axes then the Centre Point location of the Mask pattern is at [–H*tan(z)*cos(a) , –H*tan(z)*sin(a) ]. By correlating the Mask hole pattern with the detector illumination pattern its (x,y) Centre Point can be located and then the source FOV angle (z,a).

For XGIS the Mask Code pattern is a 57x57 random pattern, 50% open, 50% closed with 10x10x1 mm squares of Tungsten. The Mask Code is supported top and bottom by a 3x3 mm Aluminium alloy matrix frame (bottom reinforced with a 13x3 central cross), as shown below.

The Mask is placed on the top of a "hopper" shaped support Collimator, made of 1.0 mm Aluminium alloy side panels, strengthen by an external diamond patterned rib structure, whose interior sides are covered by a thin Tungsten shielding layer of 0.25 mm thickness. Figure 5 shows the envelope area as well as a detail of the dimensions of the frame support and the pixel size.

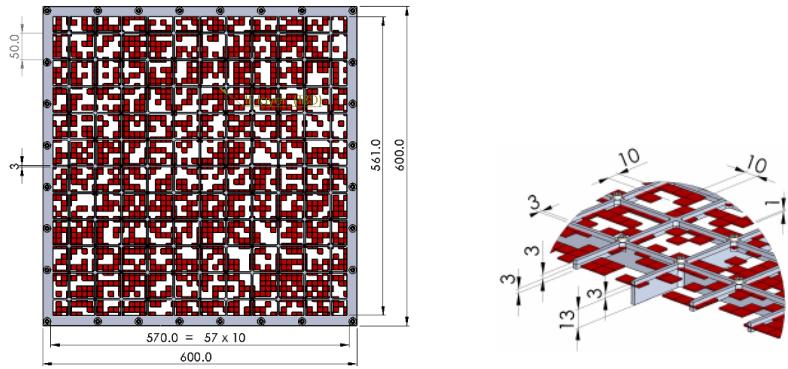

Figure 5 XGIS Coded Mask Assembly overall area (*left side*), Mask pixel size and support frame dimensions (*right side*).

### 4.3 Collimator Assembly mechanical design

The Collimator Assembly has two main objectives: to mechanically connect the Coded Mask Assembly with the detector assembly and to act as a lateral Passive Shielding for the Imaging System. The Collimator Assembly has two elements: Collimator and Passive Shielding.

The Collimator is made of Aluminium alloy 1 mm thickness and the necessary stiffeners to provide enough strength and stiffness to the sheet. The Collimator also accommodates the Passive Shielding (0.25mm W) of XGIS Imaging System.

At the same time, the Passive Shielding provides the required opacity to shield the diffuse cosmic ray background. The Passive Shielding is made of four Tungsten slabs along the Collimator height. The Collimator Assembly includes the mechanical interface with the Detector Plane, made by 16screws of M8.

The combination of the coded area with the Collimator aperture in this geometry leads to the following FOV of 78x78 deg up to energies of 150 keV for each Detector Unit, as it is depicted in the next figure (Figure 6).

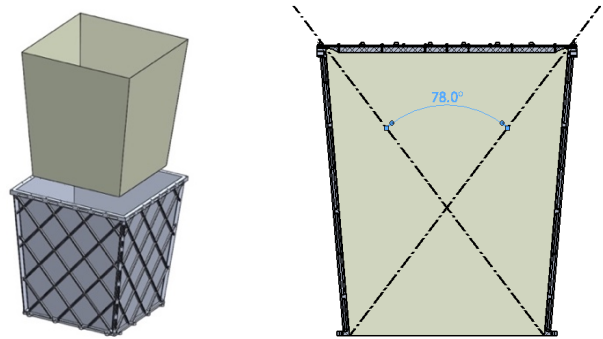

Figure 6 *Right*, XGIS Collimator Assembly exploded view showing the Passive Shielding and the Collimator element with stiffeners. Left, XGIS FOV up to 150 keV

Then the XGIS detector plane, at the bottom of each XGIS Detector Unit, detects both the low-energy radiation (up to 150 keV) passing through the XGIS Imaging System. Radiation at E > 150 keV reaches the detector from all directions because the system is "transparent" above this energy.

### 4.4 Physical properties

The Mask Code and the Passive Shielding are made of Tungsten, while the rest of the Coded Mask Assembly and the Collimator are from Aluminium alloy. Then, the overall mass budget of the XGIS Imaging System is below 30 kg, including a 30% of maturity margin.

As expected, its Centre of Gravity is placed in the XGIS central axis, with a height close to 400 mm from the Collimator Assembly bottom.

### 4.5 Thermal design

The XGIS Imaging System thermal control is fully passive. This System is in contact with the Detector Plane support frame and is the most exposed Detector Unit part to outer radiative environment (see Figure 7).

The preliminary thermal design is fully based on the mechanical design; the Collimator Aluminium alloy outer surface is coated with a chromium process (Surtec 650V), while the inner Passive Shielding surface is black painted, so that it can act as a radiative heat sink for the XGIS detector top plane.

Details of the thermal analysis of the whole XGIS system and of its imager component are reported in [3].

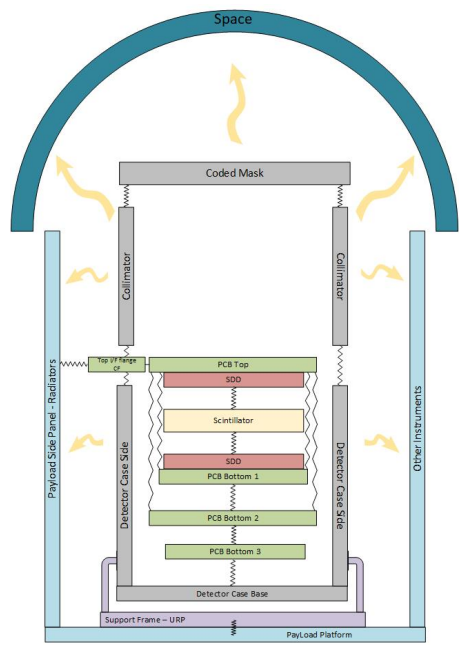

Figure 7 The thermal scheme of the XGIS Detector Unit displaying the parts considered in the preliminary design.

## 5. ANALYSES RESULTS

### 5.1 Structural Analysis

To verify the fulfilment of the mechanical requirements (stiffness, strength and stability), a set of analyses has been carried out and are described in this section. The analyses performed are:
- Normal Mode analysis
- Quasi-static
- Instability

## 5.2 Model

Main mechanical parts, which affect the fundamental frequency and control the load carrying capability, have been meshed with structural elements; these are the primary structure (Mask Support Structure and Collimator). Mask Code pixels, Passive Shielding and mechanical joint items (bolts, adhesives, miscellaneous) have been represented by means of non-structural mass elements.

As is typical in space structural analysis for this preliminary phase the FEM (Finite Element Model) is relatively simple to focus on the global behaviour. This allows to accommodate the main structure leaving the details for future phases when the design is more mature. Typically, the FEM is made up of a combination of bar and shell, but in this occasion some solid elements are also introduced to model, for the higher thickness vs width areas or significant protuberances.

Figure 8 displays a view of the Imaging System FEM.

Notice that some elements have not been modelled yet because they are not completely defined at this time, but their mass contribution has been added as a non-structural mass spread through adjacent structural elements.

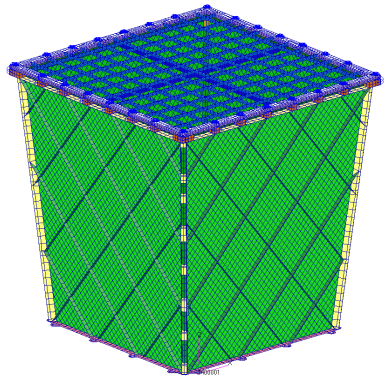

Figure 8 XGIS Imaging System Finite Element Model performed with MSC/PATRAN software.

As a reference of the model dimensions, the summary of the MSC/NASTRAN software entries count is disclosed in the following table (Table 1):

Table 1 XGIS Imaging System MSC/PATRAN software FEM entries count summary for each item.

| Items | Summary |
|---:|---|
| Grid | ~23000 nodes |
| CQUAD4 | ~12000 elements |
| CTRIA3 | ~650 elements |
| CHEXA | ~5000 elements |
| CPENTA | ~150 elements |
| RBE2 | ~50 elements |

The FEM geometry and element topology have been checked to fulfil the mesh requirements: model geometry and topology, rigid body motion strain energy, static analysis under unitary gravity load, free-free modal analysis, and free thermo-elastic deformation checks. The results of all these checks have been satisfactory. Mass and Centre of Gravity of the MSC/PATRAN software model fits with the 3D design mechanical model.

In the following analyses the unit is fixed at the interface points with its supported structure without any adjacent structure, clamped at all displacement degrees of freedom (d.o.f.) 1 to 6.

## 5.3 Results

As it is explained in Section 3, the XGIS Imaging System has been conceived to have its fundamental eigenfrequency above 60Hz, and a quasi-static load for all possible combinations with a maximum acceleration of 24 g in each direction has been considered. Stiffness, strength and deformation have been checked to fulfil the requirements and are presented in the following paragraphs.

To obtain the global stiffness of the structure (in terms of natural frequencies) a real modal analysis has been run. Table 2 lists the first eigenfrequencies and Effective Mases & Inertias associated to each one.

Table 2 XGIS Imaging System eigenfrequencies up to mode 10, including the effective mass and inertia for each one

| Mode No. | Freq. (Hz) | Modal eff. Mass [%] | | | Inertia fraction [%] | | |
|---|---|---|---|---|---|---|---|
| | | Tx | Ty | Tz | Rx | Ry | Rz |
| 1 | 59.29 | 0.0% | 0.0% | 10.9% | 2.0% | 1.4% | 0.0% |
| 2 | 82.52 | 0.0% | 0.0% | 0.0% | 0.0% | 0.0% | 0.0% |
| 3 | 87.11 | 22.8% | 0.0% | 0.0% | 0.0% | 13.4% | 7.6% |
| 4 | 87.12 | 0.0% | 22.8% | 0.0% | 12.6% | 0.0% | 4.6% |
| 5 | 98.60 | 0.0% | 0.0% | 0.0% | 0.0% | 0.0% | 0.0% |
| 6 | 115.67 | 0.5% | 0.0% | 0.0% | 0.0% | 0.1% | 0.2% |
| 7 | 115.67 | 0.0% | 0.5% | 0.0% | 0.1% | 0.0% | 0.1% |
| 8 | 117.39 | 0.0% | 0.0% | 0.0% | 0.0% | 0.0% | 0.0% |
| 9 | 128.96 | 16.7% | 7.2% | 0.0% | 10.9% | 26.6% | 1.1% |
| 10 | 128.97 | 7.2% | 16.7% | 0.0% | 25.0% | 11.6% | 11.7% |

The first global eigenfrequency is nearly 60 Hz, being a Coded Mask bending mode shape quite fixed at the Collimator upper interface flange. As can be seen in the following Figure, for the first mode, the most demanded regions are the central part of the Mask Support Structure frame. Next, in the second mode all lateral Collimator panels vibrate in consonance.

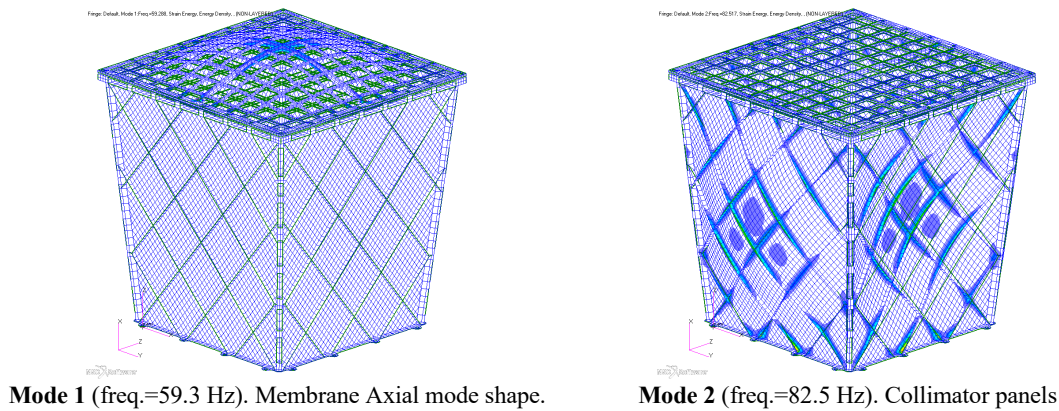

**Mode 1** (freq.=59.3 Hz). Membrane Axial mode shape.    **Mode 2** (freq.=82.5 Hz). Collimator panels

Figure 9. XGIS Imaging System first and second eigenfrequencies located at the central part of the Mask Support Structure frame and at the Collimator panels respectively.

The design limit load cases have been applied to the XGIS Imaging System to guarantee that the structure withstands this load environment without any damage. Both material strength and local/global structural instability have also been checked.

The highest stresses appear around the clamped support of the cross stiffeners on the perimeter of the square frame, and also, but at lower level, on the cross stiffeners centre due to the global grid bending. However, it is a comfortable Margin of Safety (MoS) considering a Factor of Safety (FoS) of 1.1 under yield limit for strength analysis and 1.25 under ultimate limit for instability analysis [14]. Notice that these analyses are based on the herein proposed loads (heritage), since the mission specification does not address this issue yet.

On the whole, the stresses are smaller than the material capability. The lower margin is attained in the Coded Mask Assembly at the cross stiffener to external frame connection where they present a significant reduction in height.

Collimator lateral panel stiffeners stabilize the panel skin, confining the first buckling mode into a cell defined by those lateral stiffeners. Figure 10 depicted the highest stress areas of the XGIS Imaging System.

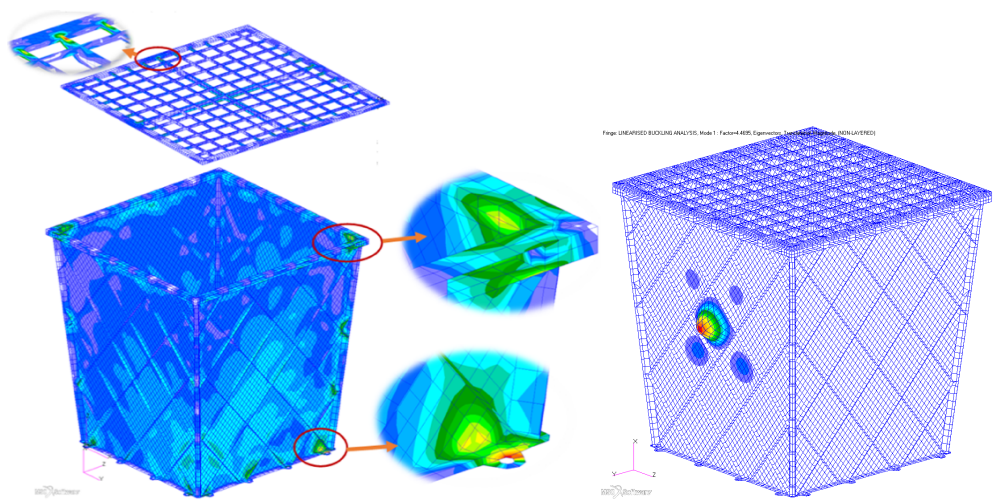

Figure 10. XGIS Imaging System load cases analysis. Red colour represents the highest stress areas

Finally, Table 3 collects the summary of the MoS results for the XGIS Imaging System strength analysis.

Table 3 XGIS Imaging System strength results summary.

|  | Quasi-static loads | | | |
|---|---|---|---|---|
|  | Strength analysis | | Instability analysis (buckling) | |
| **Imaging System part** | Max Stress | Min MoS | Buckling load factor | Min MoS |
| Collimator Assembly | 41 MPa | > 6.8 | 4.36 | > 1.4 |
| Coded Mask Assembly | 125 MPa | > 1.6 | | |

## 5.4 Thermal Analysis

## 5.5 Model

The Coded Mask Assembly and the Collimator Assembly have been simulated as part of the XGIS Detector Unit thermal model (Figure 11).

The Coded Mask Assembly covering the top surface of the Detector Unit is a complex assembly of filled and hollow square cells mounted on an Aluminium alloy frame; it has been modelled as a single shell. The open fraction of 50% is taken into account by defining a dedicated optical set of properties with 50% transmission in IR and visible bands.

The preliminary model geometry also includes the four Aluminium alloy Collimator panels reinforced by stiffness frames and covered by a Tungsten layer in the inner part. The thermo-optical properties for the outer surface of Aluminium alloy are taken from the Surtec 650V coating which has an emissivity of 0.03, while the Tungsten inner surface is painted with an emissivity of 0.8.

The base of the Collimator Assembly is bolted on the XGIS support frame directly on three sides. On the fourth side it is mounted on the cold finger flange through an insulating layer, considered as ideal in the model.

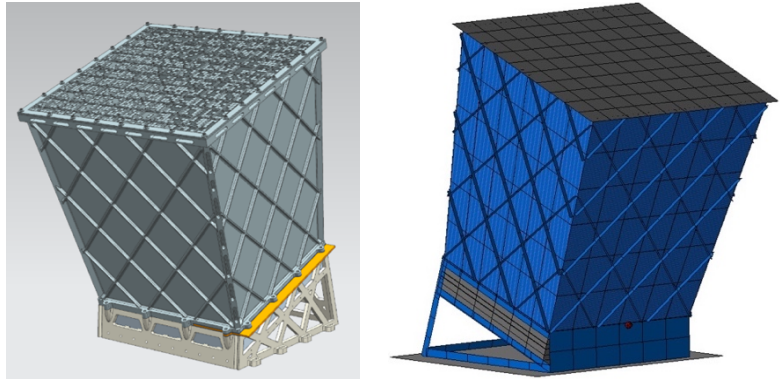

Figure 11. XGIS mechanical design (left) and thermal model geometry (right). The Coded Mask Assembly is simulated as a single square shell; the Collimator is bolted to the Mask Support Structure: on the cold finger (yellow structure) side, an insulating distancing layer is located between the interface surfaces.

## 5.6 Results

The XGIS thermal model has been run simulating the planned orbit, in order to estimate space environment effects. These have a relevant impact, in particular for the Imaging System thermal behaviour since it is the coldest part of the instrument.

The next figure (Figure 12) shows the temperature distribution of the Coded Mask Assembly and the Collimator Assembly:

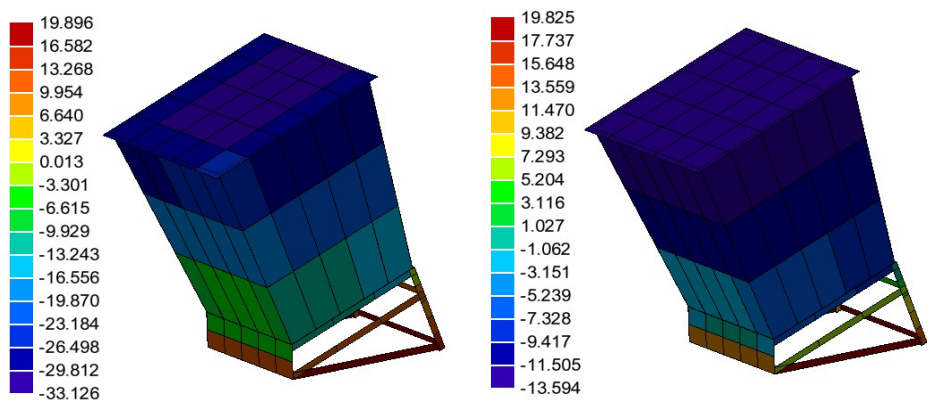

Figure 12 Temperature distribution of the Coded Mask Assembly and Collimator Assembly for cold (left) and hot (right) case analysis. The support frame is mounted on the platform with a boundary temperature of 20°C for both cases.

Likewise, Table 4 summarizes the cold and hot steady state results derived from the XGIS reduced thermal model:

Table 4 Thermal analysis output temperature data for XGIS Imaging System.

| Imaging System part | Cold Case | | | Hot Case | | |
|---|---|---|---|---|---|---|
| | Average T [°C] | Min T [°C] | Max T [°C] | Average T [°C] | Min T [°C] | Max T [°C] |
| Collimator Assembly | -20.5 | -29.9 | -6.2 | -4.1 | -8.2 | 3.3 |
| Coded Mask Assembly | -29.6 | -33.1 | -26.3 | -7.7 | -8.2 | -7.3 |

If THESEUS enters phase-B, the use of Multi-Layer Insulator (MLI) will be studied and analysed to maintain temperature gradients at levels that do not affect the imaging quality of the overall system.

## 5.7 Performance Analysis: XGIS image reconstruction and source location

The output from any XGIS observation is a list of photon events in its detector pixels which can be used to create an illumination pattern for photons in a certain energy range [15].

The simplest image of what lies in the FOV can be made by scanning an array of source locations in the XGIS FOV and making a correlation of their expected detector illumination pattern with the observed pattern. Then a linear Maximum Likelihood or Chi-squared (CHI2) regression method is used to solve for a source strength and residual detector background.

The output of this is an image of Maximum Likelihood or Signal/Noise ratio showing locations of source candidates above detector background as shown in Figure 13 and in Figure 14.

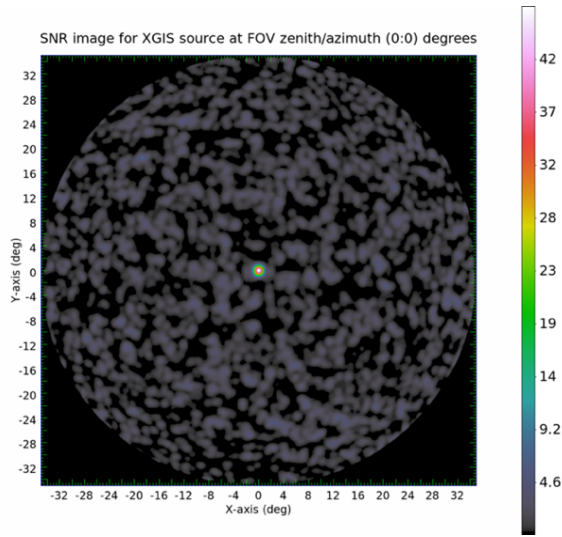

Figure 13 Fisheye SNR image reconstruction for a single on-axis source in the XGIS field of view

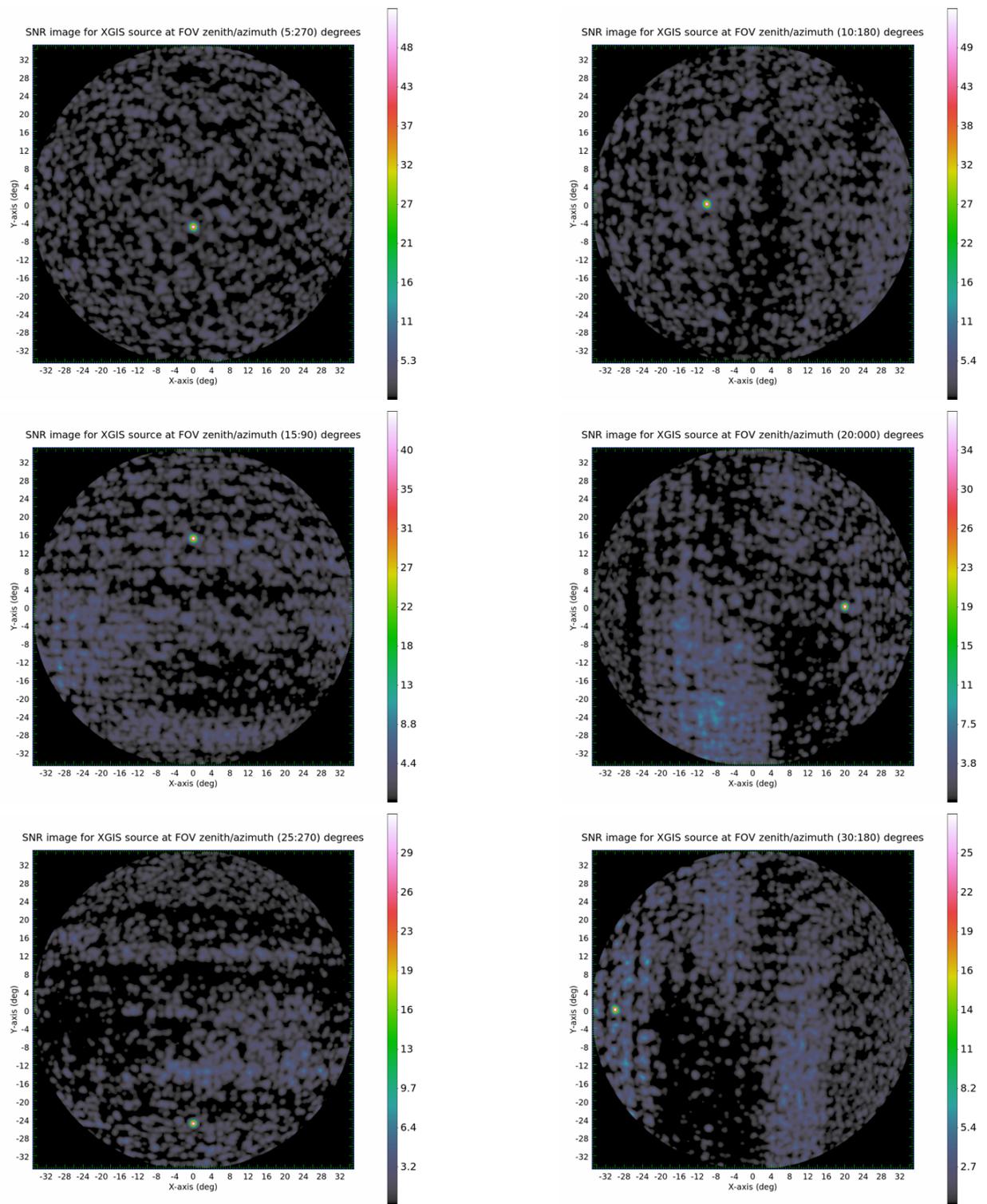

Figure 14 SNR image reconstruction for six 7.0 ph/cm$^2$ sources at (5,10,15,20,25,30º) off-axis

Figure 15 shows a SNR correlation image for five sources in the FOV at (0,5,10,20,30) degrees off-axis. Four of them stand out clearly but the fifth, at zenith/azimuth (30,180), could be interpreted as a background artefact.

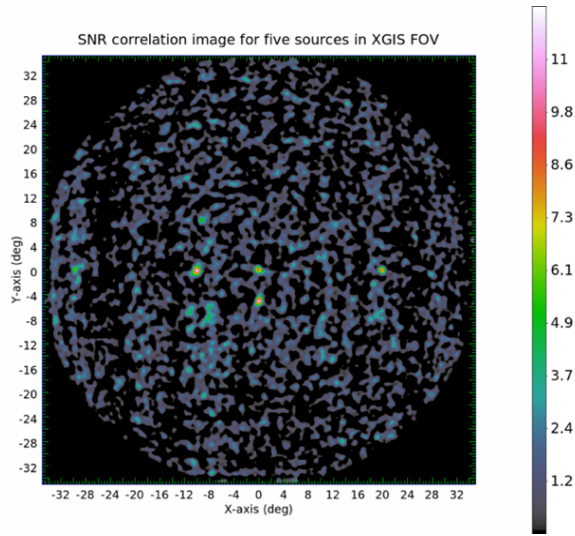

Figure 15 Fish eye SNR correlation image for five sources in field of view

Multiple sources are best located by an image reconstruction method which solves for a large number of image pixels simultaneously. This uses some smoothing or Maximum Entropy constraint, or by the method of iterative removal of sources (IROS). In this method a sequence of correlation images is reconstructed, and the strongest source location extracted; it is then appended to a source list and included in the solution so that the next correlation image is for any remaining sources.

The next Figure shows the result when the correlation image above is the input for a sequential search where four sources can be found. They are superimposed on the correlation image of remaining counts, where the fifth source, at zenith/azimuth (30°,180°), is not found because its SNR is below a certain stopping threshold.

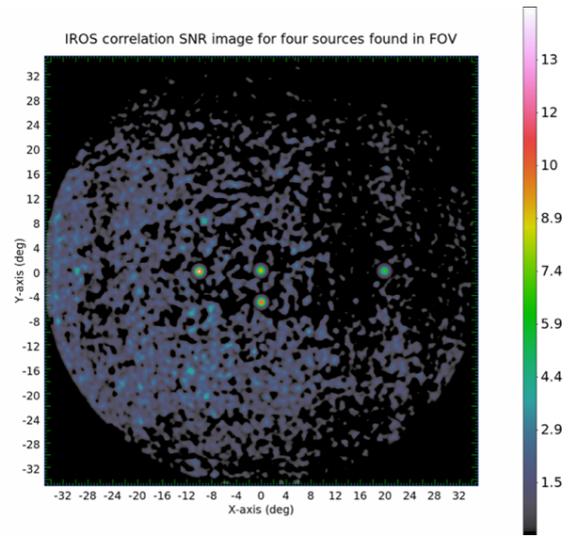

Figure 16 IROS image of four sources located and superimposed on the residue correlation image

## 6. CONCLUSIONS

- The Imaging System preliminary design of the XGIS instrument on-board the THESEUS mission, one of the M5 ESA candidates, has been presented. This design is based on signal multiplexing technique using a coded mask. Envelope and mass requirements of the Imaging System are complied with.

- Imaging System both FEM and structural analysis are included in this paper. Results fit well with the requirements assumed at this project stage. Mission stress and stiffness requirements will be considered in further steps of the project.

- Thermal model and analyses are also presented. If THESEUS enters phase-B, the use of MLI will be studied to maintain temperature gradients at levels that do not affect the system imaging quality.

- A preliminary imaging performance analysis of the XGIS Imaging System has also been included in this work.

## ACKNOWLEDGEMENTS

THESEUS/XGIS instrument is supported by: the ASI-INAF Agreement n. 2018-29-HH.0; the OHB Italia/INAF-OASBo Agreement n.2331/2020/01; the European Space Agency ESA through the M5/NPMC Programme; the AHEAD2020 project funded by the UE through H2020-INFRAIA-2018-2020; the Spanish Ministerio de Ciencia e Innovación, PID2019 109269RB-C41; the Polish National Science Centre, Project 2019/35/B/ST9/03944 and Foundation for Polish Science, Project POIR.04.04.00-00-5C65/17-00.

The authors thank all the members of the THESEUS team involved in the development of XGIS instrument.

## REFERENCES

[1] Stratta, G., et al., "THESEUS: A key space mission concept for Multi-Messenger Astrophysics", Advances in Space Research, Vol.62 (3), 662-682 (2017).

[2] Amati, L., et al., "The X/Gamma-rays Imaging Spectrometer (XGIS) on-board THESEUS: Science case, requirements, concept, and expected performances", Paper 11444-279, this Conference

[3] Labanti, C., et al, "The X/Gamma-ray Imaging Spectrometer (XGIS) on-board THESEUS: Design, main characteristics, and concept of operation", Paper 11444-303, this Conference

[4] Hutchinson, I., et al. "The soft x-ray imager on THESEUS: The transient high energy survey and early universe surveyor", Paper 11444-304, this Conference

[5] Götz, D., et al. "The Infra-red Telescope on board the THESEUS mission", Paper 11444-305, this Conference

[6] Meregheti et al., "Scientific simulations and optimization of the XGIS instrument on board THESEUS", paper 11444-276, this Conference

[7] Sánchez F., et al., "Integral Signal Multiplexing", Nuclear Instruments and Methods in Physics Research Section A, Vol. 537 (3), p. 571-580 (2005).

[8] Neubert, T., et al., "The ASIM Mission on the International Space Station", Space Science Reviews 215, Article number: 26 (2019)

[9] Soffita, P., et al., "XIPE: the X-ray Imaging Polarimetry Explorer", Experimental Astronomy, Aug. 2013

[10] Reglero, V., et al., "Low Energy Gamma Ray Imager on Minisat 01", Proceeding 2$^{nd}$ INTEGRAL, Workshop "The Transparent Universe", ESA-SP-382, p. 343-347, 1996

[11] Jeong, S., et al., "UBAT of UFFO/Lomonosov: The X-Ray Space Telescope to Observe Early Photons from Gamma-Ray Bursts", Space Science Reviews 214, Article number 16 (2018)

[12] Østgaard, N., et al., "The Modular X- and Gamma-Ray Sensor (MXGS) of the ASIM Payload on the International Space Station", Space Science Reviews volume 215, Article number: 23, 2019

[13] Neubert, T., et al, "A terrestrial gamma-ray flash and ionospheric ultraviolet emissions powered by lightning", Science 367, 183–186, 2020

[14] ECSS, ECSS-E-ST-32C Rev. 1, "Structural general requirements" (15 Nov 2008)

[15] Caroli, E., et al., "Coded aperture imaging in X- and gamma-ray astronomy", Space Science Reviews, vol. 45, no. 3-4, p. 349-403, 1987